# Understanding the growth of high-aspect-ratio grains in granular L1$_0$-FePt thin-film magnetic media

Chengchao Xu[1, 2], Bing Zhou[1, 3], Tianxiang Du[1, 3], B.S.D.Ch.S Varaprasad[1, 2], David E. Laughlin[1, 2, 3], and Jian-Gang (Jimmy) Zhu[1, 2, 3].

1) Data Storage Systems Center, Carnegie Mellon University, Pittsburgh, Pennsylvania 15213, USA
2) Electrical and Computer Engineering Department, Carnegie Mellon University, Pittsburgh, Pennsylvania 15213, USA
3) Materials Science and Engineering Department, Carnegie Mellon University, Pittsburgh, Pennsylvania 15213, USA

**Abstract**

A systematic investigation has been performed to optimize the microstructure of L1$_0$-FePt-SiO$_x$ granular thin film as recording media for heat-assisted magnetic recording. The FePt-BN nucleation layer, which is stable even at 700℃, is used to control the grain sizes and microstructure during the high-temperature processing. The study finds that films of high-aspect-ratio FePt grains with well-formed silicon oxide (SiO$_x$) grain boundaries require the grading of the deposition temperature during film growth as well as the grading of the silicon oxide concentration. Well-isolated columnar grains of L1$_0$-FePt with an average height greater than 11 nm and diameters less than 7 nm have been achieved. Transmission electron microscopy (TEM) analysis of the microstructures of samples produced under a variety of non-optimal conditions is presented to show how the microstructure of the films depends on each of the sputtering parameters.

**Introduction**

$L1_0$-ordered FePt-X granular magnetic films are currently being used as the recording media for heat-assisted magnetic recording (HAMR), where X stands for grain boundary material, such as carbon (C), silicon oxide (SiOx), and boron nitride (BN), etc. Well isolated, highly ordered and highly textured FePt grains with narrow size distribution is required for HAMR. To achieve the desired areal recording density capability (ADC), the FePt grains need to be well-separated and fully encircled by the grain-boundary materials ensuring lateral thermal isolation as well as lateral exchange decoupling between adjacent magnetic grains. As disk drive ADC continues to increase, grain size and grain-to-grain pitch distance in the media need to be scaled down accordingly while maintaining the desirable microstructure.[1] The reduction of grain size requires an increase of grain height in proportion since the recording performance has been proved to be highly correlated to the grain volume.[2] Consequently, achieving high-aspect-ratio columnar grains which form desired microstructure with thermally isolating grain boundary material becomes the key to the advancement of granular $L1_0$-FePt media.

In practice, the fabrication of granular $L1_0$-FePt media with high-aspect-ratio columnar grains and well-defined grain boundaries has remained to be challenging. Various grain boundary materials, often referred to as segregants, have been investigated over the past.[3] The FePt-carbon system[4] has been demonstrated to generate small FePt grains well encircled by amorphous carbon as grain boundaries. However, the high mobility of carbon and its high interfacial energy with FePt results in sphere-like FePt grains.[5] After the FePt-C film reaches a certain height, a second layer forms, preventing the growth of grains of high aspect ratio. Various oxide as segregants have been shown to improve the columnar growth, including $SiO_2$ [6,7], $TaO_x$ [8] and $TiO_2$ [9,10], but most of them fail to display good magnetic properties for a variety of reasons. In particular, the FePt-SiO$_2$ system has shown good columnar growth, but interconnections between neighboring FePt grains are commonly observed, which lead to a maze-like microstructure. Furthermore, the degree of L1$_0$ ordering of the FePt grains was found to be suppressed in the FePt-SiO$_2$ films,[7] resulting in low perpendicular coercivities ($H_C$). To mitigate the challenges, varying the medium compositions through its depth by grading the percentage of grain boundary materials during deposition have been employed for FePt-C and FePt-SiO$_x$ films[7,11] in attempts to create high grain aspect ratios. Moreover, the investigation of medium fabrication has expanded to bilayer structures, such as FePt-C/FePt-SiO$_x$[9] and FePt-BN/FePt-SiO$_x$[12], using either FePt-C or FePt-BN to create a growth template for improving the microstructure of the second layer FePt-SiO$_x$. Our recent studies showed that FePt-BN/FePt-SiOx bilayer system also achieved well-isolated grains with good chemical ordering.[12]

From all the previous studies cited above, what has been missing is the understanding of how the deposition conditions are affecting the resulting microstructure and how to adjust them to mitigate the challenges of obtaining high grain-aspect-ratio with well-defined grain boundaries. The study presented in this paper aims to understand the missing link through a systematic experimental study using FePt-BN/FePt-SiO$_x$ bilayer system as the example for this investigation. For the remainder of the paper, we will present an optimized fabrication method followed by the discussion of the film characteristics with controlled deviation of the deposition conditions.

**Experimental**

All the thin films samples in this study were deposited on Si (001) substrates with native oxide in an AJA sputtering system with base pressures of 2×10$^{-8}$ Torr or lower. The crystalline texture of all the samples and the averaging degree of chemical ordering of the FePt grains were examined using the standard X-ray diffraction (XRD) technique with Cu $K_\alpha$ radiation. The magnetic properties were measured using a superconducting quantum interface device vibrating sample magnetometer (SQUID-VSM) (Quantum Design MPMS3 system) with an applied magnetic field up to 7T. The microstructure of the samples was evaluated by plane-view and cross-sectional transmission electron microscopy (TEM) imaging, using bright-field TEM (BF-TEM), high-resolution TEM (HR-TEM) and scanning TEM-high angle annular dark field (STEM-HAADF) techniques using FEI Titan Themis 200. The grain size and grain center-to-center pitch distances were analyzed using the plane-view STEM-HAADF images and the image processing software (MIPAR) for accuracy. The pitch distance analysis counted the 6 nearest neighbors for each grain.

**Results and discussion**

The film stack of Ta (4 nm)/ Cr (40 nm)/ MgO (8 nm), was used for each sample in this study as underlayer. Specifically, the Cr layers were DC-sputtered at 250°C, followed by annealing at 650°C for 1 hour in vacuum to improve the (001) texture and increase the grain size. The samples were, then, cooled to about 90°C and the MgO layers were deposited using RF sputtering at 10 mTorr Ar pressure.

A thin layer of pure FePt (0.5 nm) followed by a layer of FePt-36 vol.% BN (1.6 nm) were deposited at 700°C. The FePt was deposited via DC sputtering and the BN was deposited via RF sputtering at 5 mTorr. The purpose of the initial 0.5nm pure FePt was to form nuclei with higher grain density. The 1.6-nm-thick FePt-38 vol.% BN layer was performed by the repeated deposition of [FePt-9 vol.% BN (0.22 nm, co-sputtered)/ BN (0.1 nm, RF-sputtered)] bilayer for five times. This was performed because the co-sputtering with individual FePt and BN targets can only produce FePt-9 vol.% BN, due to the significantly low sputter yield of BN relative to that of FePt. This initial [FePt(0.5 nm)/ FePt-BN(1.6 nm)] bilayer was the same for all the samples presented in this study.

The stack of five FePt-SiO$_x$ sublayers was then deposited on top of the initial FePt-BN/FePt bilayer, with various oxide compositions and at gradually decreased substrate temperatures: First, FePt-40vol.%SiO$_x$(1.6nm) was deposited followed by FePt-36vol.%SiO$_x$(1.6nm), both co-sputtered at 550°C substrate temperature. The decreasing concentration of SiO$_x$ was realized by changing the power of the SiO$_2$ targets during the deposition of each sublayer. After the deposition of the first two sublayers of FePt-SiO$_x$ layers, the substrate heater was turned off. Next, the remaining three sublayers, FePt-24% vol.SiO$_x$(1.6 nm), FePt-20% vol.SiO$_x$(1.6 nm), and FePt-18% vol.SiO$_x$(3.2 nm), were sequentially deposited while the substrate cooled from 550°C to about 430°C in the vacuum. The entire film stack is summarized in the schematic shown in Figure 1(a). Table 1 shows the target information and the sputtering conditions used in the deposition of the magnetic layers.

Figure 1(c) shows the XRD pattern of this sample. The integrated-intensity ratio of the L1$_0$ ordered super-lattice peak (001) to the fundamental peak (002) ($I_{001}/I_{002}$) is about 2.2, corresponding to an order parameter of $S$=0.72 for the L1$_0$ ordering, with the film thickness and other XRD factors considered.[13] Although the relatively sharp (002) and (001) peaks indicate the overall good texture of the FePt grains, the presence of a relatively small FePt (111) peak and the visible shoulder of the FePt (002) peak indicate the existence of small amounts of FCC phase and in-plane variants, nonetheless.

Figure 2(a) shows the plane-view BF-TEM images, in which well-separated FePt grains with well-defined narrow grain boundaries can be observed. Most of the FePt grains are fully encircled by the oxide segregant in the grain boundary areas, except for a few elongated or irregular-shaped grains that resulted from the lateral connection of neighboring grains. The very low percentage of inter-grain connections shown in this sample is insignificant contrast with that in the usual FePt-SiO$_x$ granular films, where such inter-grain connections are far more common.[6,7] The suppression of the formation of the connected grains is partially owing to the FePt-BN layer underneath the FePt-SiO$_x$ layer and to the optimization of the composition variation as well as the deposition conditions. The measured grain height is about 11.5 nm, and the average grain diameter is 7.0 nm. Thus, a grain aspect ratio (h/D) of 1.64 was achieved in this sample. The grain size distribution is shown in Figure 2(c), exhibiting a unimodal distribution.

The cross-section BF-TEM image (Figure 2(b)) shows clear columnar shaped grains, most with clear separation with grain boundaries of relatively uniform thickness through the depth. Infrequently, a lateral connection between two neighboring grains can be seen, mostly near the top of the film. Figure 1(b) shows the perpendicular and in-plane magnetic hysteresis loops, which show a perpendicular coercivity of 21 kOe and an in-plane coercivity of 5 kOe, measured at room temperature. The slightly lower perpendicular coercivity and slightly higher in-plane coercivity shown here is attributed to the imperfection of the MgO texture with relatively large distribution of (002) axis (XRD rocking curve showed FWHM > 6°) as well as the relatively high roughness of the MgO underlayer, which is quite evident in the cross-section TEM pictures.

The gradual variation of the oxide composition in the FePt-SiO$_x$ sublayers through the thickness of the film coupled with deposition temperature change are critical for obtaining the high-aspect-ratio FePt grains while maintaining good granular microstructure with well-defined grain boundaries.

While showing the above results of the optimized situation, it is also important to understand what happens if the degree of the composition variation or deposition conditions are not optimized. Below we categorize the undesired consequences into four different growth modes as the composition and deposition conditions deviate from the optimized one.

**Undesired Growth Modes**

   **1. Oxide Overage: Afloat FePt Layer**

   As shown in Figure 1(a), the FePt-SiO$_x$ sublayers have gradually decreasing SiO$_x$ compositions during the film growth. This reduction of grain boundary material as the film thickness increases is critically important for maintaining the columnar growth of the FePt grains. Let us examine the case in which we increase the SiO$_x$ concentration in the top sublayer. In this case, the film stack (Figure 3(a)) contains two sublayers of the FePt-SiO$_x$, both deposited at 550°C. The first 4.8nm-thick FePt-35 vol.%SiO$_x$ is similar to the sum of sublayers (1) to (3) in Figure 1(a). What's different is that the SiO$_x$ concentration of the top 1.6-nm sublayer was increased to 40%. Figure 3(b)-(d) shows the TEM micrographs of the film sample, both in planar and cross-sectional views. Instead of continued columnar growth, the deposition of the higher-SiO$_x$-concentration sublayer yields a dense layer of small FePt crystallites "floating" on top of SiO$_x$. This is also evident in the plane-view STEM-HAADF image with small FePt crystallites on the top. This phenomenon has been thought to be the grain height limit for FePt-SiO$_x$ media. Whereas now we know, by reducing SiO$_x$ content along with adequate deposition temperature, a greater grain aspect ratio can be achieved.

   **2. Oxide Deficiency: Bridging Connection of Neighboring FePt Grains**

   If, however, the SiO$_x$ content is not sufficient, the grain boundaries are likely to have a smaller height than that of the grains during film growth. In such cases, neighboring FePt grains could extend their growth laterally and bridge over the grain boundary (due to the lower height of the grain boundary material), forming connections between the grains, as shown in the schematic of this mechanism Figure 4(d). Figure 4(a), (b) show the TEM micrographs of a sample that contains a film stack nearly identical to the one shown in Figure 1(a), except that both sublayers (4) and (5) have 18 vol.% $SiO_x$ (which are packed into the 4.8-nm sublayer in Figure 4(c)). The FePt bridging over silicon oxide grain boundaries can be clearly seen, showing the impact of the segregant deficiency. The portion of the FePt that bridges over the $SiO_x$ is found to be ordered in the right direction if the connected grains sit on the same MgO grain. However, if the connected grains are located on different sides of the MgO grain boundaries, and hence have different in-plane orientations, the bridging-over portions of FePt are not likely to be textured and ordered in the desired way, similar to the formation of multi-variant grains across the MgO grain boundaries.[14] This observation indicates that it is important to ensure that the grain boundary material and FePt grains always maintain the same height during the growth process to avoid either the re-nucleation of FePt crystallites due to grain boundary material coverage, or the bridging between neighboring FePt grains over the grain boundary due to the deficiency of the grain boundary material. One may also notice that the microstructure is quite sensitive to the FePt-$SiO_x$ composition through the thickness. A couple of volumetric percentage changes can strongly influence the resulting microstructure.

**3. Over Heating: Neighboring Grain Connection via Grain Boundary Penetration**

In this section, we discuss the effect of deposition temperature. A film sample with the stack (shown in Figure 5(a)) including the sublayers (1) to (4) shown in Figure 1(a) were fabricated with similar deposition conditions, except that the sublayer (4) was deposited at 650 °C. Figure 5 shows the TEM study of the sample. The cross-section TEM images show that many neighboring FePt grains are connected laterally in the FePt-$SiO_x$ layer while they are still separated by the BN grain boundaries in the bottom FePt-BN layer. Such lateral inter-granular connections create elongated worm-like grains and thus lead to a maze-like microstructure in the plane-view image, which is similar to those of FePt-SiO$_x$ layer directly deposited on MgO at similar deposition temperatures.

Figure 5(b) clearly indicates that at 650 °C, the silicon oxide grain boundary is not effective to confine the FePt grains, allowing their lateral expansion and coalescence across the grain boundaries to minimize the interfacial energy. Whereas the boron nitride grain boundary material is still stable in this case. Since the sputtered silicon oxide is less dense and should has a much lower glass transition temperature, 650 °C can be close to the glass transition point. The silicon oxide in between adjacent grains at this temperature could become easily deformable (plastic) and be "squeezed" away by the lateral growth of the sandwiching FePt grains. We also suspect the boron nitride grain boundaries are partial crystalline since they are able to endure well this relatively high temperature environment.

### 4. Under Heating: Disrupted Growth

Figure 6 shows the TEM images of a sample with its stack and deposition conditions included in the figure. The deposition temperatures of the FePt-SiO$_x$ sublayers have been lowered compared to the optimal deposition conditions shown in Figure 1(a). Specifically, it also adopted a decreasing SiO$_x$ concentration, but the entire FePt-SiO$_x$ layer was deposited at temperatures lower than 550 °C. In other words, the cooling program was brough forward to start from the first FePt-SiO$_x$ sublayer, as shown in Figure 6(a). The cross-section image, Figure 6(b), shows broken or disrupted grains through the depth, especially near the top of the stack. If we examine the cross-section image, an apparent layer of oxide covering the top is not observed in this case. Hence, such discontinuous and disrupted growth is evidently the result of the temperature lower than the optimal one. The real cause should be the relatively low surface mobilities of silicon oxide at such temperatures. As the film grow thicker, the area packing fraction of FePt increases. The incoming silicon-oxide adatoms with lower surface mobility, due to the low substrate temperature, are more likely to "stick" on the FePt surface, thus disrupting the grain growth. Therefore, the lower-than-optimum deposition temperature always results in broken grains due to disrupted growth.

**Summary**

In this study, the growth of granular $L1_0$-FePt thin film media is systematically investigated, via the introduction of a FePt-BN and FePt-SiO$_x$ bilayer system. An optimized FePt-BN bottom layer provided a stable granular template for the growth of FePt-SiO$_x$. To attain desired microstructure with good perpendicular coercivity, the SiO$_x$ composition and deposition conditions are varied. The growth of the FePt-SiO$_x$ layer was divided into five sublayers and the composition and substrate temperature were varied for each sublayer deposition. With optimization, we were able to achieve high-aspect-ratio columnar $L1_0$-FePt grains, with a height of 11.5 nm at 7.0 nm diameter, encircled by well-defined grain boundaries. One of the keys to achieving the optimized growth is the adequately gradual reduction of the silicon oxide concentration along the bottom-up direction through the thickness of the FePt-SiO$_x$ layer. This composition variation must be coupled with an appropriate gradual reduction of the substrate temperature during the deposition process as well.

Throughout this study, we were also able to characterize the undesired growth modes and associate the modes with composition and sputtering conditions as summarized in Figure 7. If silicon oxide material is excessive after filling all the gaps between adjacent grains, small FePt crystallites form over the excessive oxide layer, as shown in Figure 7(d). This situation is referred to as the afloat second layer. On the other hand, silicon oxide deficiency also yields uneven heights among the FePt grains and grain boundaries during film growth. This often leads to lateral bridging of FePt between neighboring grains over the surface of oxide grain boundaries, as shown in Figure 7(c). If the substrate temperature during deposition of FePt-$SiO_x$ is higher than the glass transition temperature of sputtered silicon oxide, the oxide no longer can separate FePt grains well and neighboring FePt grains tend to connect laterally through the entire thickness of the layer, as shown in Figure 7(a). Finally, if the substrate temperature is lower than optimum, discontinuous grain growth occurs due to the low mobility of the oxide, as shown in Figure 7(b).

**Acknowledgements**

This research was funded in part by the Data Storage Systems Center at Carnegie Mellon University and all its industrial sponsors and by the Kavcic-Moura Fund at Carnegie Mellon University. The authors acknowledge the use of the Materials Characterization Facility at Carnegie Mellon University supported by Grant No.MCF-677785. The authors would also like to thank Bo Yuan Yang for his help.

**Data availability**

The data that supports the findings of this study are available within the article.

**Author Declarations**

The authors have no conflicts to disclose

| Targets | Purity | Sputter condition | Power(W) | Deposition rates (Å/s) |
|---|---|---|---|---|
| $Fe_{55}Pt_{45}$ | 99.99% | DC-sputter, 5 mTorr | 20 | 0.14 |
| BN | 99.5% | RF-sputter, 5 mTorr | 150 | 0.015 |
| $SiO_2$ | 99.999% | RF-sputter, 5 mTorr | 100 | 0.1 |

Table I. Information and deposition conditions of the targets used in the magnetic layer.

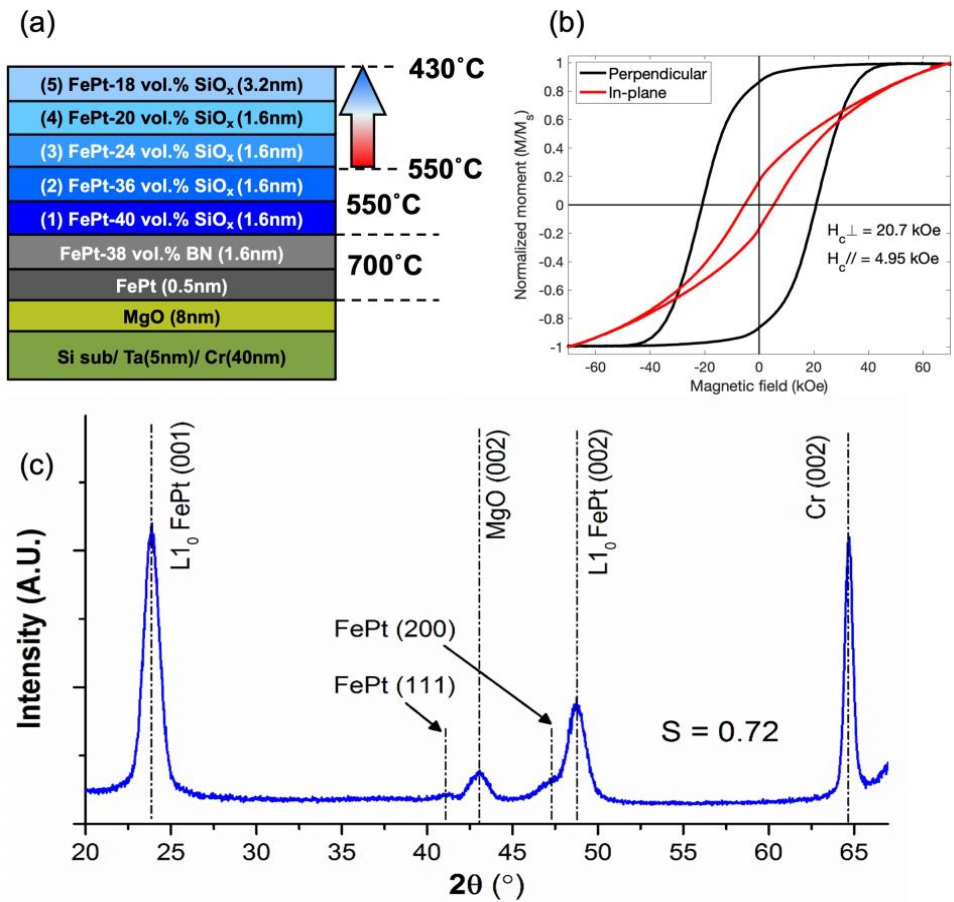

FIG. 1. Film stack and characterizations of the sample fabricated with the optimized compositions and deposition conditions: (a) Schematics of film stack with Ta/Cr/MgO underlayers, FePt/FePt-BN template layers, and the optimized compositions and substrate temperatures for each FePt-$SiO_x$ sublayer, (b) Magnetic hysteresis loop, (c) Out-of-plane X-ray diffraction pattern.

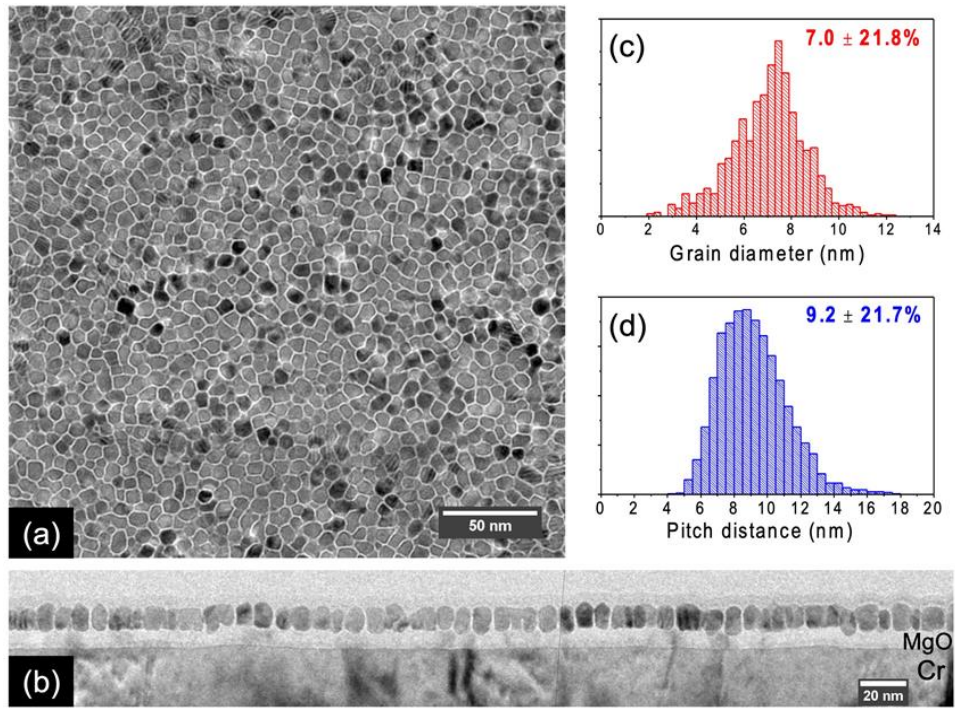

FIG. 2. TEM micrograph of the sample fabricated with the optimized compositions and deposition conditions: (a) Plane-view BF-TEM image, (b) Cross-sectional BF-TEM image, with a Cr capping layer intended to reveal the height of grain boundary materials. (c) Grain size distribution; (d) Grain center-to-center pitch distance distribution.

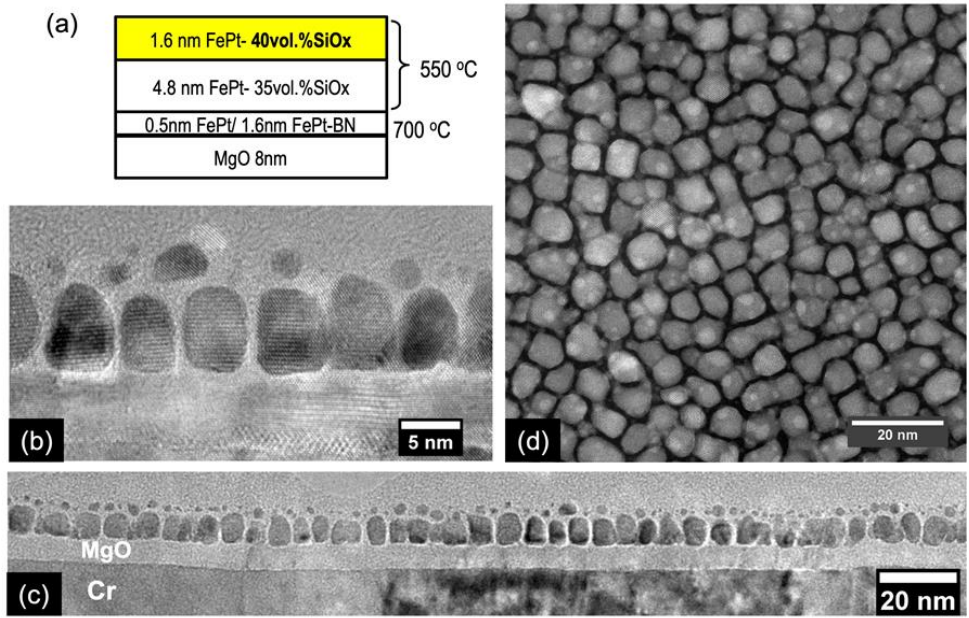

FIG. 3. TEM study of the sample with excessive SiO$_x$ relative to the optimized compositions: (a) film stack; (b) cross-section view HR-TEM image and (c) cross-sectional BF-TEM images; (d) plane-view STEM-HAADF image of this sample. TEM micrographs reveal a layer of small FePt grains, which re-nucleated over the oxide grain boundary material that covered the bigger grains. This is referred to as the afloat second-layer small grains in this paper.

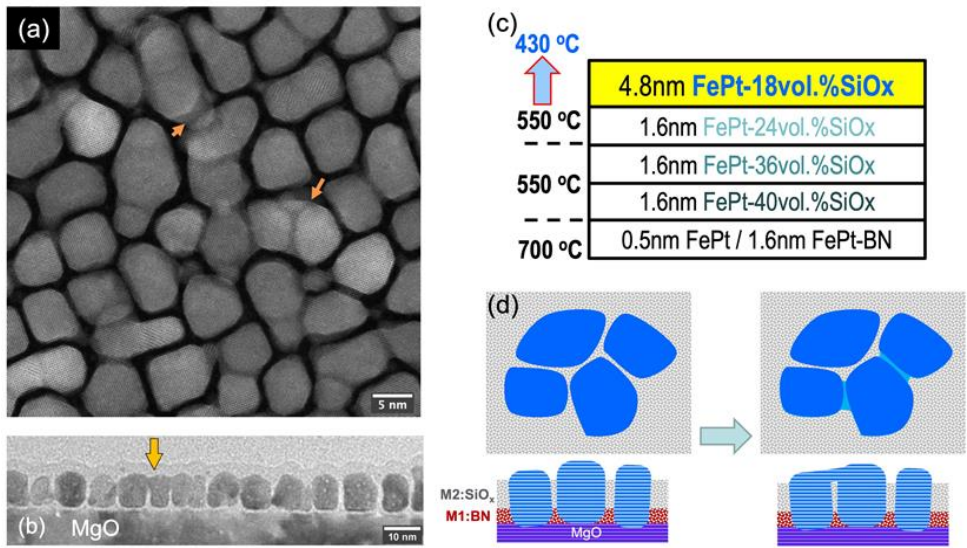

FIG. 4. TEM study of the sample with a controlled reduction of SiO$_x$ concentration in the top FePt-SiO$_x$ sublayers relative to the optimal one: (a) Plane-view STEM-HAADF image, (b) Cross-section view BF-TEM image with a Cr capping layer to reveal the height of grain boundaries, and (c) its detailed film stack. (d) Evolution of the microstructure when the grain boundaries are lower than the FePt grains. The portion of FePt that bridges over the oxide appears brighter in HAADF imaging, hence the existed grain boundaries that are covered are evident.

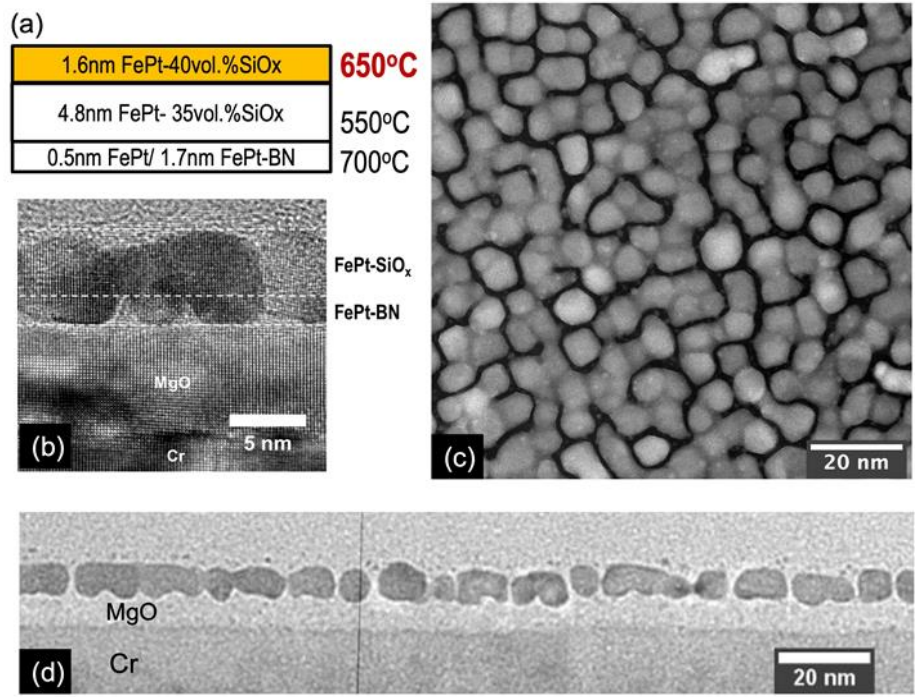

FIG. 5 TEM study of a sample with the last FePt-SiO$_x$ sublayer deposited at a higher temperature (650 °C) than the optimized sample: (a) Film stack; (b) HRTEM image of the structure of a worm-shape inter-grain connection. The lower parts of the grains are still separated by boron nitride. (c) Plane-view STEM-HAADF image showing a maze-like microstructure; (d) Cross-sectional BF-TEM image.

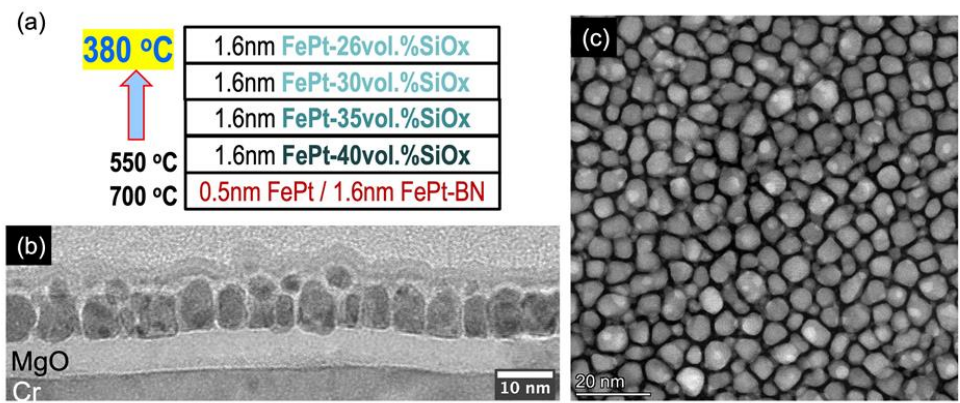

FIG. 6. TEM study of the sample with the FePt-SiO$_x$ layer deposited at lower temperatures: (a) Film stack, (b) Cross-sectional BF-TEM image of the sample, with a Cr capping layer to indicate the grain boundary materials; (c) Plane-view STEM-HAADF image shows scattered broken grains that overlap with each other.

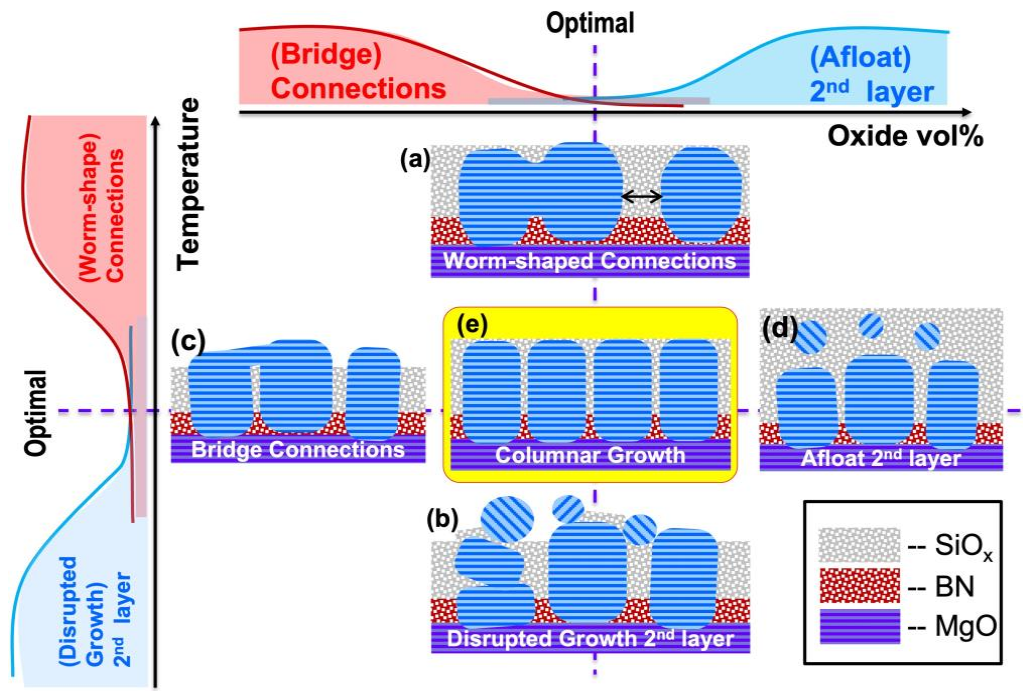

FIG. 7. Schematics of the four typical undesired growth modes that are commonly observed in the TEM study of the FePt-BN/FePt-SiO$_x$ bilayer medium samples. Two axes represent the probabilities of the growth modes associated with the deviation of two critical parameters from the optimum: deposition temperature and volume fraction of silicon oxide segragant. (a) Worm-shape inter-grain connections are caused by over-heating. (b) Disrupted grain growth arises from lower temperatures. (c) Bridge inter-grain connections resulted from segregant deficiency, and (d) Afloat second layer caused by excessive segregant, are associated with mismatched height among FePt grains and oxide grain boundaries. (e) Columnar growth achieved with optimized deposition conditions.